\documentclass[10pt,conference]{IEEEtran}
\usepackage{epsfig}
\usepackage{verbatim}
\usepackage{balance}
\usepackage{amsmath}
\usepackage{cite}
\usepackage{float}
\usepackage{algorithm}
\usepackage{amssymb}
\usepackage[english]{babel}
\usepackage{array}
\usepackage{comment}
\usepackage{multirow}
\usepackage{enumerate}
\usepackage{color}
\usepackage{xcolor}
\usepackage{listings}
\usepackage{graphicx}
\usepackage{epstopdf}
\usepackage{graphics}
\usepackage{hyperref}
\usepackage{enumitem}
\usepackage{textcomp}
\usepackage{tikz}
\usepackage{textcomp}
\usepackage[doipre={DOI:~}]{uri}
\usepackage{lipsum}
\usepackage{gensymb}
\usepackage{siunitx}
\usepackage{caption}
\usepackage[acronym]{glossaries}
\newacronym{lom}{LoM}{Logic-on-Memory}
\newacronym{mol}{MoL}{Memory-on-Logic}
\newacronym{sip}{SiP}{System-in-Package}
\newacronym{hpc}{HPC}{High-Performance Computing}

\usepackage{tikz}

\usepackage{tabularx}
\newenvironment{conditions*}
{\par\vspace{\abovedisplayskip}\noindent
	\tabularx{\columnwidth}{>{$}l<{$} @{${}={}$} >{\raggedright\arraybackslash}X}}
{\endtabularx\par\vspace{\belowdisplayskip}}

\def\BibTeX{{\rm B\kern-.05em{\sc i\kern-.025em b}\kern-.08em
    T\kern-.1667em\lower.7ex\hbox{E}\kern-.125emX}}

\usepackage{fancyhdr}

\fancypagestyle{firstpage}{
    \fancyhf{} 
    \fancyfoot[C]{%
        \colorbox{gray!20}{%
            \parbox{0.95\textwidth}{%
                \small
                \textcopyright 2025 IEEE. Personal use of this material is permitted.  Permission from IEEE must be obtained for all other uses, in any current or future media, including reprinting/republishing this material for advertising or promotional purposes, creating new collective works, for resale or redistribution to servers or lists, or reuse of any copyrighted component of this work in other works.\\
                Accepted for publication at the 51st IEEE European Solid-State Electronics Research Conference (ESSERC 2025).
            }%
        }%
    }
}

\title{Thermal Implications of Non-Uniform Power in BSPDN-Enabled 2.5D/3D Chiplet-based Systems-in-Package using Nanosheet Technology \vspace{-1.0ex}}

\author{
\IEEEauthorblockN{
Yukai Chen\IEEEauthorrefmark{1},
Massimiliano Di Todaro\IEEEauthorrefmark{1}\IEEEauthorrefmark{3},
Bjorn Vermeersch\IEEEauthorrefmark{1},
Herman Oprins\IEEEauthorrefmark{1},
Daniele Jahier Pagliari\IEEEauthorrefmark{3},\\
Julien Ryckaert\IEEEauthorrefmark{1}
Dwaipayan Biswas\IEEEauthorrefmark{1}
James Myers\IEEEauthorrefmark{2}
}
\IEEEauthorblockA{\IEEEauthorrefmark{1}IMEC, Leuven 3000, Belgium\\}
\IEEEauthorblockA{\IEEEauthorrefmark{2}IMEC, Cambridge CB1 2JD, United Kingdom\\}
\IEEEauthorblockA{\IEEEauthorrefmark{3}DAUIN, Politecnico di Torino, Turin 10129, Italy\\
Email: yukai.chen@imec.be}
\vspace{-6ex}
}

\begin{document}

\maketitle
\thispagestyle{firstpage}  

\IEEEaftertitletext{\vspace{-2mm}}

\begin{abstract}
Advances in nanosheet technologies have significantly increased power densities, exacerbating thermal management challenges in 2.5D/3D chiplet-based Systems-in-Package (SiP). While traditional thermal analyses often employ uniform power maps to simplify computational complexity, this practice neglects localized heating effects, leading to inaccuracies in thermal estimations, especially when comparing power delivery networks (PDN) in 3D integration. This work examines the thermal impact of non-uniform power distributions on SiPs utilizing frontside (FSPDN) and backside (BSPDN) power delivery approaches. Using high-resolution thermal simulations with non-uniform power maps at resolutions down to \qty{5}{\micro\meter}, we demonstrate that uniform power assumptions substantially underestimate peak temperatures and fail to reveal critical thermal differences between BSPDN and FSPDN configurations in 3D scenarios. Our results highlight that BSPDN configurations in 3D, although beneficial in simplified uniform scenarios, exhibit pronounced thermal penalties under realistic, localized workloads due to limited lateral heat spreading. These findings emphasize the necessity of adopting fine-grained, workload-aware power maps in early-stage thermal modeling to enable accurate PDN assessment and informed thermal-aware design decisions in advanced nanosheet-based 3D SiP.
\end{abstract}

\section{Introduction and Motivation}
\label{sec:intro}
Semiconductor scaling into the nanosheet era has significantly increased computing core density, causing elevated power densities and heightened thermal management challenges in high-performance computing (HPC) systems. Chiplet-based Systems-in-Package (SiP), integrating multiple dies either horizontally (2.5D integration) or vertically (3D integration), compound these thermal challenges due to their compact and complex multi-layered structures. Consequently, effective heat dissipation has emerged as a critical design requirement, influencing system performance and reliability.

Backside Power Delivery Networks (BSPDN) have recently emerged to improve power integrity by delivering power directly from the wafer backside~\cite{beyne2023nano}. By relocating power rails to the backside, BSPDN reduces interconnect resistance, lowers IR drop, and enhances power efficiency, freeing frontside metal layers for optimized signal routing. However, BSPDN implementation involves extreme silicon substrate thinning, substantially affecting thermal pathways. Previous studies indicated that substrate thinning inherent to BSPDN significantly raises hotspot temperatures due to diminished lateral heat spreading. Although backside metal layers partially mitigate this effect, elevated peak temperatures remain compared to conventional Frontside Power Delivery Networks (FSPDN)~\cite{oprins2022package}. While BSPDN’s thermal impact has been evaluated in 2D System-on-Chip (SoC) scenarios, these findings do not generalize to 3D integrations, where vertical stacking introduces fundamentally different thermal dynamics.

Traditional early-stage thermal analyses frequently rely on simplified uniform power maps to reduce computational complexity. However, uniform power assumptions inadequately represent realistic chip operations, where computationally intensive workloads produce localized hotspots.
Recent literature demonstrates significant underestimations of peak temperatures using uniform maps, particularly emphasizing these inaccuracies in 2D SoC with BSPDN~\cite{lofrano2024block}.
Likewise, workload-based non-uniform power maps reveal that BSPDN exhibits hotspot temperatures up to \qty{14}{\degree} higher than comparable FSPDN designs in the 2D HPC SoC scenarios~\cite{bjorn2024}. Such findings highlight the necessity of detailed, realistic non-uniform power distributions for accurate thermal modeling.

Nevertheless, most existing studies remain limited to 2D SoC with a single active power layer, leaving comprehensive thermal analyses for complex, vertically integrated 2.5D/3D chiplet-based SiPs largely unexplored. Addressing this critical gap, our study systematically investigates the thermal implications of employing fine-grained non-uniform power maps in BSPDN-enabled 2.5D/3D SiPs, explicitly focusing on 3D integration scenarios. 
Our study rigorously compares the thermal behavior of BSPDN and FSPDN under realistic workload-driven and synthetic power distributions, revealing critical hotspots and thermal gradients previously masked by uniform power assumptions in 3D chiplet integration.

\begin{figure*}[t]
	\begin{center}
		\includegraphics[width=0.93\linewidth]{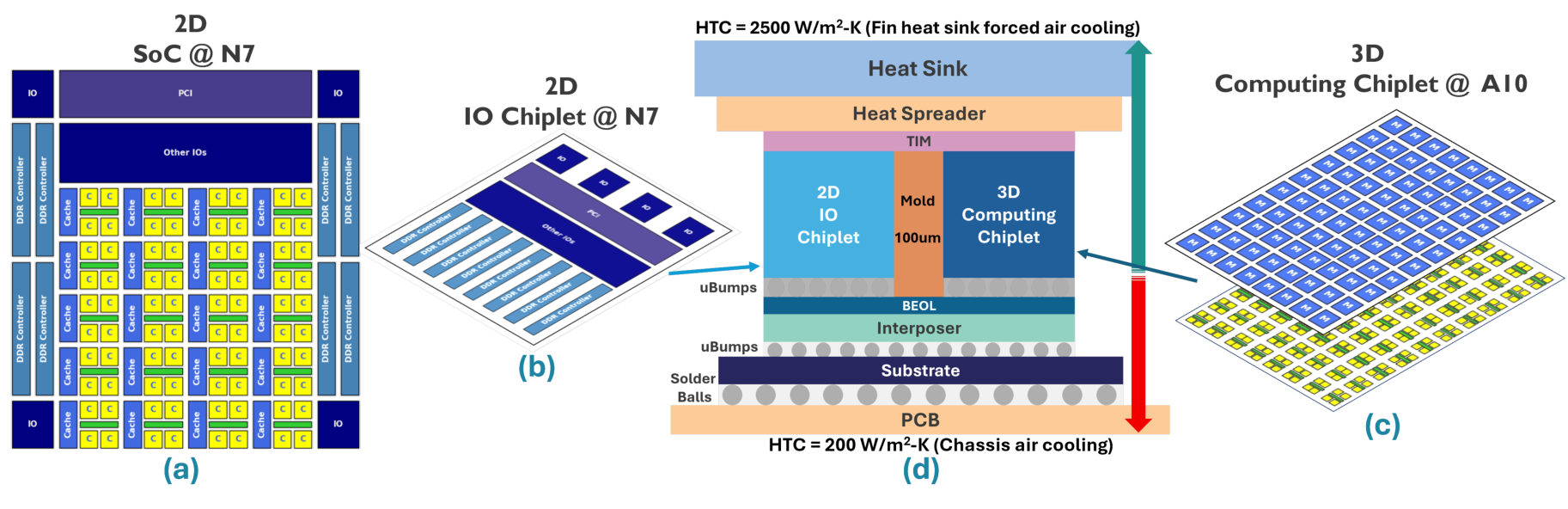}
            \vspace{-0.3cm}
		\caption{(a) The original 2D HPC SoC implemented in N7; (b) Floorplan of the 2D I/O chiplet disaggregated from the SoC implemented in N7; (c) Floorplan of the 3D computing chiplet implemented in A10; (d) 2.5D chiplet-based SiP thermal model.}
		\label{fig:system} 
	\end{center}
        \vspace{-0.8cm}
\end{figure*}

In 3D design with uniform power maps, BSPDN may appear thermally superior to FSPDN, which is contradictory to the 2D analyses. However, our high-resolution thermal simulation reveals the opposite under realistic power distributions. 
By adopting non-uniform power maps with resolutions as fine as \qty{5}{\micro\meter}, we identify thermal penalties inherent to BSPDN configurations. The spatial granularity of power profiles fundamentally influences hotspot formation and thermal gradients, directly affecting the accuracy of thermal estimations and interpretations of PDN effectiveness.
This study highlights that finer-grained non-uniform power maps are essential for accurate thermal modeling, thereby guiding the design and selection of effective power delivery and thermal management strategies in advanced nanosheet-based 2.5D/3D SiPs.

\section{Chiplet-based SiP and Thermal Modeling}
\label{sec:back}
To accurately evaluate the thermal characteristics of chiplet-based SiP, we consider a representative HPC server SoC originally implemented in N7 FinFET node, referencing a contemporary industrial design \cite{james2022future}. 
Figure~\ref{fig:system}.(a) shows the mock-up floorplan of this 2D SoC design. Leveraging the disaggregation methodology outlined in~\cite{chen2024thermal}, peripheral functional blocks, including PCI interfaces, I/O controllers, and DDR memory controllers, are grouped into a 2D IO chiplet fabricated in N7 technology for cost efficiency and moderate performance requirements (Figure~\ref{fig:system}.(b)). Compute-intensive components, such as cores, network-on-chip (NoC), and system-level caches, are integrated into a 3D computing chiplet using advanced A10 nanosheet technology for higher performance and density (Figure~\ref{fig:system}.(c)). This computing chiplet employs a 3D Memory-on-Logic (MoL) integration scheme, stacking system-level cache memory directly atop a logic die containing cores and NoC. The work~\cite{kumar2024thermal} presents a thermal analysis based on a similar chiplet-based design, whereas it does not consider the BSPDN and non-uniform power maps.

Figure~\ref{fig:system}.(d) illustrates the resulting 2.5D SiP thermal model. The IO and computing chiplets connect through a passive silicon interposer chosen for its simplicity and cost-effectiveness, containing no active power-dissipating elements. The narrow inter-chiplet gap (\qty{100}{\micro\meter}) filled with a low thermal conductivity mold compound (\qty[per-mode=symbol]{3}{\watt\per\meter\per\kelvin}) provides mechanical support.
Two thermal dissipation paths are modeled. The primary thermal path is through the package's topside (green arrow in Figure~\ref{fig:system}.(d)), modeling a forced-air multi-fin heatsink with a high heat transfer coefficient (HTC = \qty[per-mode=symbol]{2500}{\watt\per\meter\squared\per\kelvin}). and a secondary thermal path through interposer and PCB, representing typical chassis-air cooling conditions (HTC = \qty[per-mode=symbol]{200}{\watt\per\meter\squared\per\kelvin}). Table~\ref{table:twopathes} summarizes dimensions, thicknesses, and thermal conductivities for all stacking layers involved in both thermal dissipation paths.

\begin{table}[!htbp]
\centering
\caption{Physical parameters of stacking layers.}\label{table:twopathes}
\begin{tabular}{|c|c|c|c|}
\hline
\begin{tabular}[c]{@{}c@{}}Stacking\\ Layers\end{tabular}           & \begin{tabular}[c]{@{}c@{}}Dimension\\ (mm)\end{tabular} & \begin{tabular}[c]{@{}c@{}}Thickness\\ (um)\end{tabular} & \begin{tabular}[c]{@{}c@{}}Thermal \\ Conductivity \\ (W/mK)\end{tabular} \\ \hline
Heat Sink                                                           & 100x100                                                  & 3000                                                     & 400                                                                       \\ \hline
Heat Spreader                                                       & 40x40                                                    & 5000                                                     & 400                                                                       \\ \hline
TIM                                                                 & 32.9x22.6                                                & 250                                                      & 30                                                                        \\ \hline
\begin{tabular}[c]{@{}c@{}}IO Chiplet\\ uBumps\end{tabular}         & 19.2x22.6                                                & 10                                                       & 3.5                                                                       \\ \hline
\begin{tabular}[c]{@{}c@{}}Computing Chiplet \\ uBumps\end{tabular} & 13.6x19.6                                                & 10                                                       & 3.5                                                                       \\ \hline
Interposer BEOL                                                     & 32.9x22.6                                                & 5                                                        & 1.2                                                                       \\ \hline
Interposer                                                          & 32.9x22.6                                                & 50                                                       & 140                                                                       \\ \hline
uBumps                                                              & 32.9x22.6                                                & 100                                                      & 6.0                                                                       \\ \hline
Substrate                                                           & 50x50                                                    & 300                                                      & 0.6                                                                       \\ \hline
Solder Balls                                                        & 50x50                                                    & 100                                                      & 8.0                                                                       \\ \hline
PCB                                                                 & 100x100                                                  & 800                                                      & 5.0                                                                       \\ \hline
\end{tabular}
\vspace{-0.45cm}
\end{table}

\begin{figure}[!htbp]
       \vspace{-0.5cm}
	\begin{center}
		\includegraphics[width=0.95\linewidth]{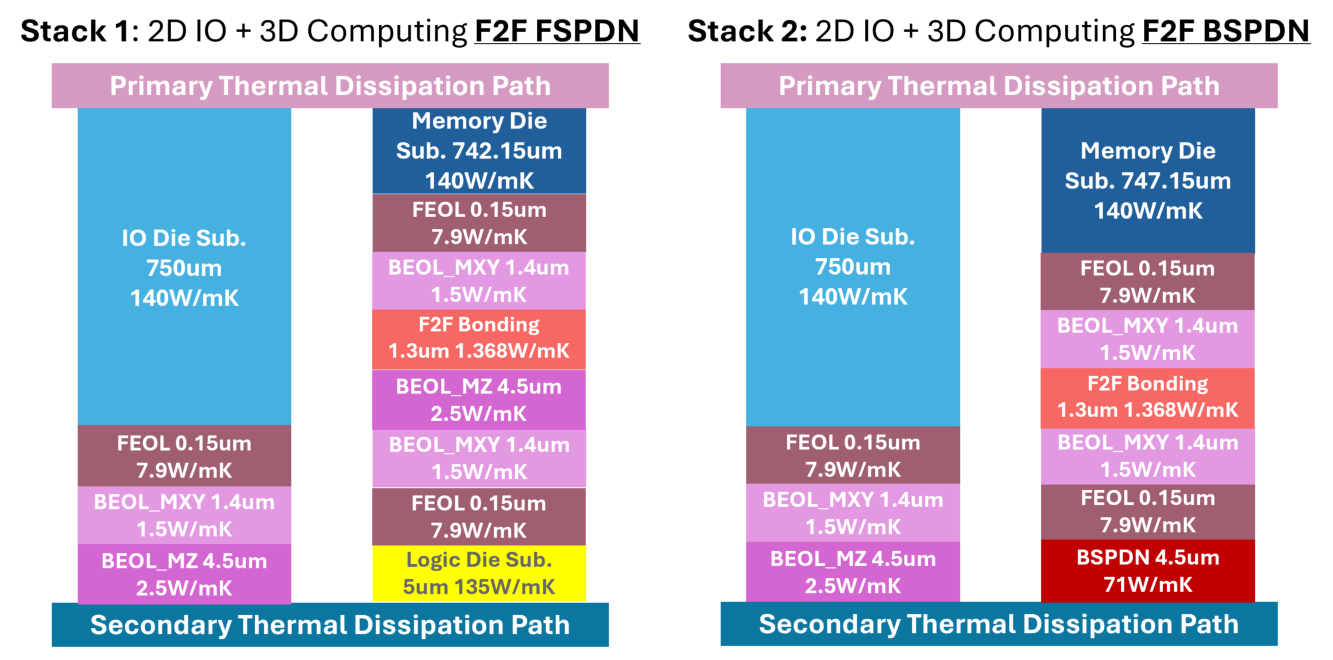}
        \vspace{-0.3cm}
		\caption{Cross sections of two chiplets (not to scale)}
		\label{fig:detail} 
        \vspace{-0.8cm}
	\end{center}
\end{figure}

Our thermal analysis specifically targets the 3D computing chiplet (\qty{13.6}{mm}×\qty{19.6}{mm}), fabricated in A10 nanosheet technology, resulting in higher power densities compared to the thermally benign 2D IO chiplet (\qty{19.2}{mm}×\qty{22.6}{mm}). Consequently, the IO chiplet primarily acts as an effective thermal spreader within the package. 
We focus our comparative thermal evaluation on two distinct power delivery network (PDN) architectures in the 3D computing chiplet: FSPDN and BSPDN. Both configurations employ face-to-face (F2F) hybrid bonding for vertical interconnections. Figure~\ref{fig:detail} details cross-sectional views of both stacks, including relevant thicknesses and thermal conductivities of stacking layers.

In stack 1 (FSPDN), power delivery is facilitated through a dedicated BEOL\_MZ metallization stack layer, distributing power to both the logic die and the vertically integrated memory die. This configuration necessitates an ultra-thin silicon substrate (\qty{5}{\micro\meter} thick) within the logic die,  accommodating nTSVs for vertical interconnections. Conversely, stack 2 (BSPDN) fundamentally alters the power delivery pathway by directly supplying power through the backside of the logic die, eliminating the BEOL\_MZ layer between logic and memory dies. The logic die's thinner silicon substrate is removed due to the backside contact implementation, and a BSPDN layer is directly underneath the logic die's FEOL layer with a lower thermal conductivity (\qty[per-mode=symbol]{71}{\watt\per\meter\per\kelvin}) compared to the thinner logic die's silicon substrate (\qty[per-mode=symbol]{135}{\watt\per\meter\per\kelvin}) in stack 1. Transitioning from stack 1 (FSPDN) to stack 2 (BSPDN) introduces two key thermal implications: 
(a) {\bf Improved vertical heat conduction} between the logic and memory dies due to the removal of the intermediate BEOL\_MZ layer, thereby reducing thermal resistance at the memory-logic bonding interface;
(b) {\bf Degraded lateral heat spreading} beneath the FEOL as the high-conductivity silicon substrate is replaced by a lower-conductivity BSPDN layer, which restricts lateral heat dissipation beneath the FEOL, exacerbating localized hotspot temperatures. 
This fundamental trade-off highlights the complexity and importance of accurate thermal modeling in guiding PDN choices for advanced 3D SiP designs.

\section{Results and Discussion}
\label{sec:results}
We evaluated the thermal impact of power map granularity and PDN architectures by conducting high-resolution thermal simulations using a hierarchical global-local methodology based on HotSpot~\cite{zhang2015hotspot}. The ambient temperature is set at \qty{25}{\degreeCelsius}.

\begin{figure}[t]
	\begin{center}
		\includegraphics[width=0.9\linewidth]{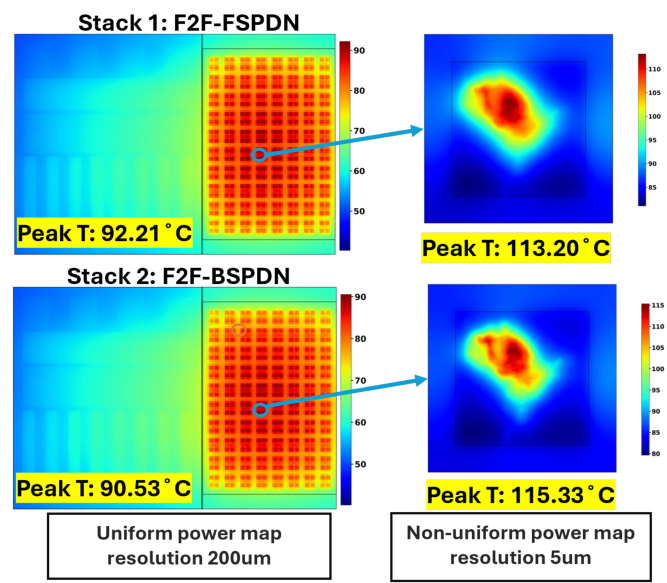}
            \vspace{-0.2cm}
		\caption{Thermal maps from uniform power and fine-grained Dhrystone workload-driven power map.}
		\label{fig:results1} 
	\end{center}
\vspace{-0.8cm}
\end{figure}

We initially performed baseline simulations employing uniform power maps at a coarse~\qty{200}{\micro\meter} resolution for each core within the 3D computing chiplet’s logic die. 
Under these uniform conditions, as shown in Figure~\ref{fig:results1} left side, BSPDN appears to yield lower peak temperatures than FSPDN, primarily due to the reduced vertical thermal resistance between the memory and logic dies achieved by removing the BEOL\_MZ layer in stack 2 (BSPDN). However, this advantage arises solely under idealized uniform power conditions, which do not realistically represent actual workloads.

\begin{figure*}[t]
	\begin{center}
		\includegraphics[width=0.9\linewidth]{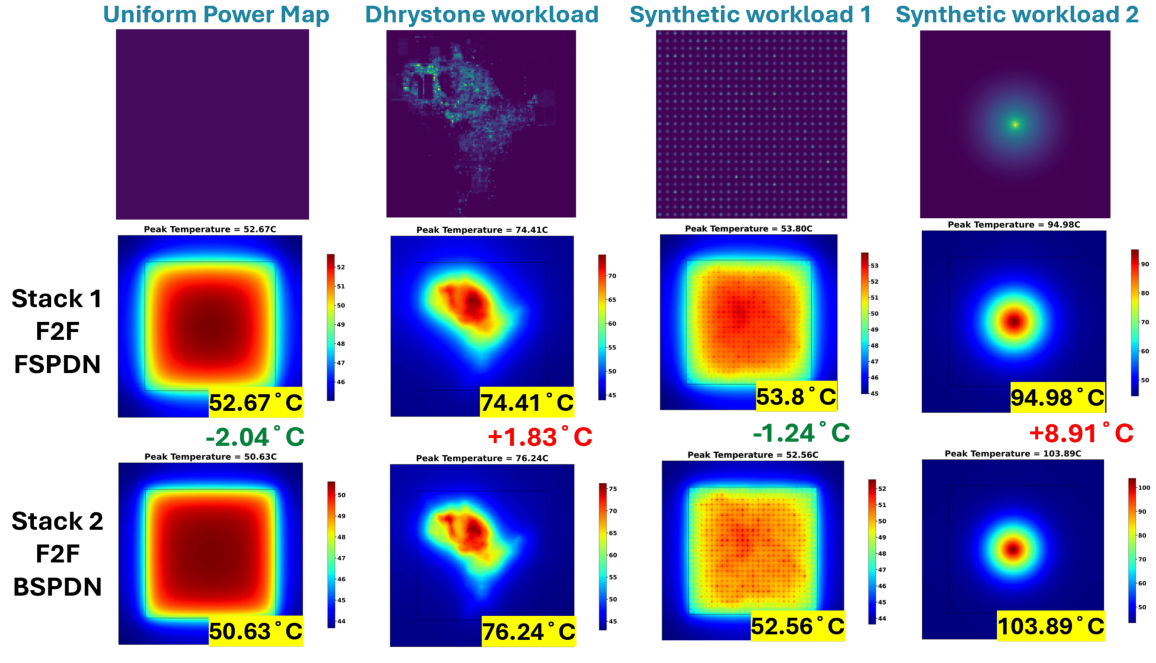}
            \vspace{-0.25cm}
		\caption{Four power maps used in single-core thermal analysis: (1) Uniform distribution, (2) Realistic workload non-uniform map, (3) Synthetic 1: clustered hotspots, (4) Synthetic 2: center-focused concentration. All cases have equal total power.}
	\label{fig:results2} 
\vspace{-0.85cm}
	\end{center}
\end{figure*}

To accurately capture realistic conditions, we selected the hottest core and applied a \qty{5}{\micro\meter} resolution workload-annotated non-uniform power map using a global-local refinement flow. This methodology first applies a coarse-grained whole-chip simulation to extract boundary thermal conditions, followed by localized refinement incorporating interpolation, filtering and upscaling to preserve spatial detail. The resulting thermal map accurately captures core-level gradients while maintaining global context.
Our results demonstrate that realistic, localized workloads significantly increase peak temperatures and fundamentally reverse the thermal trend observed under uniform power maps: BSPDN, which appears superior under uniform assumptions, consistently shows higher peak temperatures than FSPDN under realistic conditions. This reversal, extending and intensifying earlier 2D BSPDN observations~\cite{lofrano2024block}, highlights the critical importance of spatial power modeling in evaluating PDN thermal performance in 3D chiplet systems.

The reversal of thermal ranking between BSPDN and FSPDN configurations in realistic scenarios is primarily attributed to the limited lateral heat spreading capability beneath the FEOL in BSPDN implementations. 
In FSPDN, the presence of a thin silicon substrate beneath the FEOL ensures effective lateral heat spreading, moderating local hotspot temperatures. Conversely, the BSPDN architecture replaces this high-conductivity substrate with a lower-conductivity BSPDN layer, severely restricting lateral heat diffusion.
When power is spatially concentrated, this limitation becomes critical. Heat fails to diffuse laterally and instead accumulates beneath the hotspot in FEOL, intensifying local temperature rises. Although BSPDN benefits from improved vertical conduction due to a thinner bonding interface, the dominant heat escape path in 3D stacks remains through the top heatsink. Without effective in-plane diffusion, the vertical path becomes overwhelmed. 
This explains why BSPDN performs worse under realistic non-uniform power conditions and why its thermal behavior improves under uniform scenarios, allowing the heat to return more evenly through the heatsink.

To systematically explore how spatial power distribution influences thermal behavior, we conducted controlled single-core experiments using the same 3D stacking (Figure~\ref{fig:detail}), eliminating thermal interference from neighboring cores. Four power maps shown in Figure~\ref{fig:results2} with identical total power were evaluated:
1) {\bf Uniform}—power is evenly distributed across the entire core, reflecting a common early-stage modeling assumption; 
2) {\bf Realistic workload non-uniform}—extracted from workload-annotated post-PnR simulation, capturing localized activity patterns; 
3) {\bf Clustered synthetic}—the core is divided into multiple 25×25 sub-blocks, each forming a center-weighted micro-hotspot with randomized placement, emulating spatially distributed but localized compute regions; 
4) {\bf Center-focused synthetic}—high-power pixels are aggregated at the core center, creating a dominant hotspot to stress thermal limits under extreme concentration.

These four maps span a controlled spectrum from uniform to highly localized power, enabling systematic analysis of how spatial shape and location of power influence thermal gradients across different PDNs. The results confirm that spatial power concentration directly amplifies BSPDN’s thermal disadvantage.
As the power distribution becomes more aggregated, BSPDN’s peak temperature increases significantly due to the restricted lateral thermal diffusion in the low-conductivity BSPDN layer. FSPDN also exhibits peak temperature increases, but to a lesser extent, thanks to its silicon substrate’s superior ability to dissipate heat laterally. The key insight is that the hotspot size and spatial concentration determine the extent to which the material beneath the FEOL can support lateral spreading. BSPDN’s thermal penalty becomes pronounced when heat is confined to a limited area, overwhelming the vertical escape path to the heatsink.

Coarse-grained power assessments can underestimate peak temperatures and misrepresent PDN thermal behavior, which may lead to poor design choices. As power density increases in the nanosheet era and 3D integration becomes more complex, using detailed, high-resolution thermal modeling with workload-driven power maps is crucial for accurate early-stage thermal evaluations and informed PDN decisions.

\section{Conclusion}
\label{sec:concl}

Our investigation of non-uniform, fine-grained power maps for BSPDN in 3D chiplet-based SiPs uncovers an often overlooked thermal penalty: when workloads are localized, BSPDN’s constrained in-plane heat conduction can elevate hotspot temperatures beyond those in frontside-powered counterparts. This outcome contrasts sharply with the more benign predictions from uniform power assumptions. Our high-resolution thermal simulations enable the early identification of thermally critical regions, guiding effective PDN design by revealing subtle in-plane heat-spreading limitations. By eschewing coarse power assumptions and embracing fine-grained analysis, designers can more accurately gauge how localized workloads intensify hotspot formation. These insights underscore the urgency of co-optimizing power delivery and cooling strategies for nanosheet-based 2.5D/3D SiPs.


\bibliographystyle{IEEEtran}
\bibliography{reference}

\end{document}